\documentclass[12pt]{article}

\usepackage{times}
\usepackage{epsfig,amsmath}
\usepackage{subfigure}
\usepackage{graphicx}
\usepackage{color}
\usepackage{epstopdf}

\newenvironment{sciabstract}{%
\begin{quote} \bf}
{\end{quote}}

\newcounter{lastnote}

\title{Heralding Quantum Entanglement between Two Room-Temperature Atomic Ensembles}

\author
{Hang Li,$^{1,2}$ Jian-Peng Dou,$^{1,2}$ Xiao-Ling Pang,$^{1,2}$ Tian-Huai Yang,$^{1,2}$\\
Chao-Ni Zhang,$^{1,2}$ Yuan Chen,$^{1,3}$ Jia-Ming Li,$^{4, \ast}$ Ian A. Walmsley,$^{6,7}$\\
Xian-Min Jin$^{1,2,\ast}$\\
\\
\normalsize{$^1$Center for Integrated Quantum Information Technologies (IQIT), School of Physics}\\ 
\normalsize{and Astronomy and State Key Laboratory of Advanced Optical Communication Systems}\\
\normalsize{and Networks, Shanghai Jiao Tong University, Shanghai 200240, China}\\
\normalsize{$^2$CAS Center for Excellence and Synergetic Innovation Center in Quantum Information and}\\
\normalsize{Quantum Physics, University of Science and Technology of China, Hefei, Anhui 230026, China}\\
\normalsize{$^3$Institute for Quantum Science and Engineering and Department of Physics,}\\
\normalsize{Southern University of Science and Technology, Shenzhen 518055, China}\\
\normalsize{$^4$School of Physics and Astronomy, Shanghai Jiao Tong University, Shanghai 200240, China}\\
\normalsize{$^6$Clarendon Laboratory, University of Oxford, Parks Road, Oxford OX1 3PU, United Kingdom}\\
\normalsize{$^7$Blackett Laboratory, Imperial College London, London SW7 2AZ, United Kingdom}\\
\normalsize{$^\ast$E-mail: lijm@sjtu.edu.cn}\\
\normalsize{$^\ast$E-mail: xianmin.jin@sjtu.edu.cn}\\
}

\date{}

\begin{document}
\baselineskip24pt

\maketitle

\begin{sciabstract}
Establishing quantum entanglement between individual nodes is crucial for building large-scale quantum networks, enabling secure quantum communication, distributed quantum computing, enhanced quantum metrology and fundamental tests of quantum mechanics. However, the shared entanglements have been merely observed in either extremely low-temperature or well-isolated systems, which limits the quantum networks for the real-life applications. Here, we report the realization of heralding quantum entanglement between two atomic ensembles at room temperature, where each of them contains billions of motional atoms. By measuring the mapped-out entangled state with quantum interference, concurrence and correlation, we strongly verify the existence of a single excitation delocalized in two atomic ensembles. Remarkably, the heralded quantum entanglement of atomic ensembles can be operated with the feature of delay-choice, which illustrates the essentiality of the built-in quantum memory. The demonstrated building block paves the way for constructing quantum networks and distributing entanglement across multiple remote nodes at ambient conditions.
\end{sciabstract}

\subsection*{Introduction.}
The development of quantum mechanics has built strong foundations for quantum entanglement in fundamental principles and experiments. Establishing a heralded quantum entanglement between two macroscopic objects is not only of prime importance to the verification of quantum theories\cite{EPR_1935,RMP_QE}, but also a critical ability for constructing large-scale quantum networks\cite{Kimble_internet}. As the conceptual graph shows in Fig. 1a, the heralded entanglement of individual quantum nodes together with the built-in quantum memory, used to store quantum states, is an essential ingredient for building quantum networks. Together with the local operations and classical communication, quantum resources can be disseminated among the whole quantum networks, which have wide applications for quantum information processing \cite{nature_DLCZ,RMP_QR, Briegel_repeaters,QIPC_2005,nodes_QNet,YAC_teleprt}, quantum computation\cite{Copsey_computation} and quantum metrology\cite{Giovannetti_metrology,telescope_QR, clock_QNetworks}. 

So far, several seminal experimental achievements of heralding entanglement between two individual nodes have been accomplished in various systems, such as cold atomic ensembles\cite{Chou_cold}, trapped ions\cite{Moehring_ions}, solid-state crystals doped with rare-earth ions\cite{Gisin_crystals} and macroscopic diamonds\cite{science_diamond}. Furthermore, much experimental efforts have been paid to build scalable quantum repeaters based on heralded entanglement in cold atomic ensembles\cite{nodes_QNet, repeater_BDCZ}. The achieved heralded quantum entanglements between two quantum nodes, however, have to be prepared and detected in the systems that are maintained at extremely low temperature and well isolated with environment. These rigorous conditions for being avoid of strong decoherence effects limit the efficient physical scalability for quantum networks.

Apart from successful sharing quantum entanglement, quantum memories, as the stationary nodes of quantum networks\cite{optical_Qmemory}, should satisfy some key features to efficiently deliver the promised quantum advantages, including the capacities of operating at room temperature, low noise level and large time bandwidth product (the storage lifetime of entanglement divided by pulse duration). However, these capacities have been proven not compatible to each other in practice and very challenging to achieve at the same time\cite{Collision_decoherence, Raman_decoherence}. Until recently, the Duan-Lukin-Cirac-Zoller protocol operating at far off-resonance configuration has been found capable of accessing all the capacities simultaneously, especially the intrinsic low-noise mechanism\cite{FORD, npj_FORD, Hybrid_memory}. 

Here, we experimentally realize the heralding quantum entanglement between two atomic ensembles and achieve quantum network nodes simultaneously at low-noise, broadband and room-temperature regime. Billions of motional atoms collectively carry the entangled state and are separated by two centimeter-size glass cells. By mapping the stored state to photonic state, we are able to verify the heralded quantum entanglement rigorously, and more remarkably, to test the delay-choice gedanken experiment with built-in quantum memories. 

\subsection*{Experimental implement and results.}
The time sequence and energy levels for the generation and verification of the heralded entanglement are illustrated in Fig. 1b. For establishing entanglement between the two atomic ensembles (labels L, R), the optical pump pulse is split into two equal parts by controlling the incident polarization to symmetrically excite the L and R ensembles. Due to the sufficiently weak intensity of each pulse, the probability of simultaneously generating two excitations resulting from spontaneous Raman scattering in both ensembles is extremely low\cite{nature_DLCZ, Duan_PRA2002}. No matter which atomic ensemble has been excited, there will be a single collective excitation, called spin wave, shared by billions of motional atoms\cite{WStateAtoms}. Accompanying with the generation of the atomic spin wave, a correlated Stokes photon will be scattered from the ensemble with different polarization from the optical pump light. If we can erase the ``which-way" information (scattering from L or R ensemble), a spin wave will be heralded and delocalized in the two atomic ensembles. As shown in Fig. 2, we use the half-wave plate and polarizer to mix the polarization information to realize the indistinguishable detection of the Stokes photon, the resulted joint state can be written as\cite{Chou_cold}  
\begin{equation}\label{eq1}
\left | \Psi _{L,R} \right \rangle=\frac{1}{\sqrt{2}}\left (\left | 1 \right \rangle_{L}\left | 0 \right \rangle_{R}\pm e^{i\varphi  _{S}}\left | 0 \right \rangle_{L}\left | 1 \right \rangle_{R}  \right )
\end{equation}  
where $\left | 1 \right \rangle_{L,R}, \left | 0 \right \rangle_{L,R}$ represent whether there is a spin wave in L or R ensemble, $\varphi  _{S}$ is the phase difference before detecting of the Stokes photon, and $\pm$ depends on which detector receive the Stokes photon.

For verifying the existence of the heralded entanglement, we need to reconstruct the density matrix $\rho_{L,R}$ for the entangled state (1). However, the direct measurement of spin wave is an unaccessible task. Alternatively, the quantum coherence between spin waves can be transformed into the interference of anti-Stokes photon, and measured by single-photon detectors. We apply another optical probe pulse for mapping the delocalized spin wave into an anti-Stokes photon (as Figure 1b shows) after a storage time of $100 ns$. Similar with equation (1), the entangled state of spin wave distributed by two ensembles will be transformed into the entangled state of two spatial modes (L and R) for anti-Stokes photon, i.e. $\left | \Psi _{L,R}^{AS} \right \rangle=\frac{1}{\sqrt{2}}\left (\left | 1 \right \rangle_{L}^{AS}\left | 0 \right \rangle_{R}^{AS}\pm e^{i\varphi  _{S}+\varphi  _{AS}}\left | 0 \right \rangle_{L}^{AS}\left | 1 \right \rangle_{R}^{AS}  \right )$, where $\left | 1 \right \rangle_{L,R}^{AS}, \left | 0 \right \rangle_{L,R}^{AS}$ have the same meaning as equation (1), $\varphi  _{AS}$ is the phase difference for the anti-Stokes photon. In other words, the entanglement of photon retrieved from the ensembles, represented by state $\left | \Psi _{L,R}^{AS} \right \rangle$, is also heralded by the detection of a single Stokes photon, whose density matrix $\rho_{L,R}^{AS}$ can be reconstructed by the measured statistics of correlated photons. 

The experimental realization of above scheme in room-temperature atomic ensembles is organized in Fig. 2. The scattered Stokes and anti-Stokes photons with orthogonal polarizations are combined in a polarizing beam splitter, which forms a Mach-Zehnder interferometer together with the part of separating the control light. Then, the correlated  scattering photons pair is redirected to individual measurement blocks according to their different time sequences. It is worth noting that the phase of the interferometer must be stabilized, which is vital to observe the interference phenomenon of the heralded anti-Stokes photon, i.e. the phase $\varphi  _{S}$ and $\varphi  _{AS}$ must be kept constant. Therefore, an auxiliary field with feedback control has been directed into the interferometer for phase locking (see Methods). In addition, the adopted far off-resonance configuration is insensitive to the Doppler effects of motional atoms meanwhile endows the broadband feature, which allows the detections of the heralded entangled photon modes at high data rate.

Before verifying the genuine entanglement between the two atomic modes, we firstly evaluate the coherence of the two modes of atomic entangled state, which can be inferred from the interference of the anti-Stokes photon. Due to that the single spin wave is distributed between L and R ensembles, the retrieved anti-Stokes photon is also delocalized in two spatial modes with different polarizations combined by the polarizing beamsplitter in Fig. 2. By adding an extra Pancharatnam-Berry's phase\cite{science_diamond,Berry_phase,Langford_Ph.D_2007} (tuning the phase item $\varphi  _{AS}$, see Methods) between the different polarization modes, we can observe the interference of anti-Stokes photon after the projection measurement in the two output ports $D_{3}$ and $D_{4}$. The interference results $N_{\pm}$ with phase variations (the coincidence counts of $D_{1}$ and $D_{3,4}$) are shown in Fig. 3a. The experimentally measured results of $N_{\pm}$ can be fitted by sinusoidal oscillations, whose theoretical forms are proportional to $sin^{2}[\left (\varphi _{S}+\varphi _{AS}+ \pi \pm \pi \right )/2]$\cite{science_diamond}. The high fringe visibility of anti-Stokes photon between the two spatial modes implies that the quantum coherence between the L and R ensembles can be well preserved in a certain storage time.

To quantitatively evaluate the entanglement between L and R ensemble, we map it into the entanglement of anti-Stokes photon between two spatial modes and measure its concurrence (a monotonic measure of entanglement, which is positive for entangled state and zero for separable state). Due to that this transformation is local operation, it does not increase entanglement amount. Therefore, the concurrence of the photonic entanglement sets a lower bound for the atomic entanglement\cite{Chou_cold}. The density matrix of the entanglement of a single anti-Stokes photon distributed in two spatial modes, represented by $\rho_{L,R}^{AS}= \left | \Psi _{L,R}^{AS} \right \rangle \left \langle \Psi _{L,R}^{AS} \right |$, can be expressed as a matrix in the representation of photonic Fock state basis. The correlation of the concurrences for the density matrix $\rho_{L,R}^{AS}$ and $\rho_{L,R}$ can be read as\cite{concurrence, C_decoherence}
\begin{equation}\label{eq1}
C_{a}\geq C_{p}=2\,max(0, \left | d \right |-\sqrt{p_{00}p_{11}})      
\end{equation}
where $C_{a}, C_{p}$ are the concurrence of the entanglement of atomic ensembles and anti-Stokes photon respectively, $d$ is the off-diagonal coherence of $\rho_{L,R}^{AS}$, $p_{ij}$ is the heralded probability of registering $i\in \left \{0,1  \right \}$ photon in the L mode and $j\in\left \{0,1  \right \}$ photon in the R mode. 

The density matrix of the heralded entanglement of anti-Stokes modes $\rho_{L,R}^{AS}$ can be deduced from the measurement results of the correlated photons statistics, whose specific form is shown in Fig. 3b. The off-diagonal item $d$, standing for the coherence of two anti-Stokes modes, can be evaluated by the visibility $V$ of interference fringe in Fig. 3a, i.e. $d=V(p_{01}+p_{10})/2$. From the diagram illustration of $\rho_{L,R}^{AS}$, we can see that the vacuum component occupies the most proportion resulting from the weak excitation rate of scattering photons pairs (including the limited retrieval efficiency of the anti-Stokes photon), the propagation loss of  photons and the limited efficiency of detecting. According to equation (2), the concurrence $C_{p}$ for $\rho_{L,R}^{AS}$ is estimated to be $(4.5\pm0.3)\times 10^{-3}$, which well exceeds zero with 15 standard deviations, approaching the maximum value of the concurrence can reach ($C_{max}=p_{01}+p_{10}=6.6\times 10^{-3}$ with the ideal conditions of $V=1$ and $p_{11}=0$). 

 From another prospective shown in Fig. 4a, if we subtract the vacuum component of $\rho_{L,R}^{AS}$, the joint entangled state of the Stokes and anti-Stokes photons becomes
\begin{equation}\label{eq1}
\left | \Psi _{S,AS} \right \rangle=\frac{1}{\sqrt{2}}\left (\left | H \right \rangle_{S}\left | H \right \rangle_{AS}+\left | V \right \rangle_{S}\left | V\right \rangle_{AS}  \right )
\end{equation} 
where $\left | H \right \rangle_{S,AS}, \left | V \right \rangle_{S,AS}$ represent the orthogonal polarization modes of the Stokes and anti-Stokes photons. We compensate the extra Pancharatnam-Berry's phase of $\varphi  _{S}$ or $\varphi  _{AS}$ to make $\varphi  _{S}+\varphi  _{AS}=0$, forming the maximal entangled state. By performing the quantum state tomography measurement of two-qubits\cite{tomo}, the reconstructed state of our experimental state is shown in Fig. 4b. The concurrence of the reconstructed state is 0.88, and the fidelity between the experimental state and the Bell state of the equation $(3)$ is 0.92, which reflects the existence of the entanglement between the two ensembles in another way. In addition, we have measured the delocalized feature of the joint entangled state of the Stokes and anti-Stokes photons through testing the Clauser-Horne-Shimony-Holt (CHSH)-type inequality\cite{CHSH}. The correlation function results in different measurement settings are shown in Fig. 4c. The obtained $S$ value is up to $2.48\pm0.03$, with a violation of the CHSH inequality $S \leq 2$ by 16 standard deviations.

The heralded quantum entanglement together with the built-in quantum memories can display the realization of actively delaying the choice of measurement for being adapted to the future quantum information processing\cite{delay_choice,DC_science07,DC_science12}. The tomography results in Fig. 3b and 4b show strong evidences that the pair of correlated Stokes and anti-Stokes photons has the property of entanglement with memory function. We extend the delay of the projection measurement of Stokes photon about $160 ns$ and sweep the storage time to mimic the choice of measurement, which aims at testing the concurrence of our heralded entanglement. As the space-time diagram shows in Fig. 5a, the space-like separation between the Stokes and anti-Stokes detection events with not overlapping forward light cones in Fig. 5a shows that there were no causality in the delayed entanglement detections. Therefore, we should achieve similar testing results about the heralded quantum entanglement in different choices of delay time. There are three situations about the detection orders: measuring the Stokes photon first, measuring both two photons simultaneously and measuring the anti-Stokes photon first. Under the three different measuring orders, we test the quantification of the heralded entanglement between the two ensembles. As shown in Fig. 5b,c, the visibilities and concurrences with different delay choices are nearly the same with acceptable measuring errors. However, the similar testing results of the three situations imply totally different physical meanings, which are illustrated in Fig. 5a.

\subsection*{Discussion and Conclusion.}

In conclusion, we have reported the experimental realization of heralding quantum entanglement among billions of motional atoms separated by two glass cells, at low-noise, broadband and room-temperature regime. The achieved low-noise level delivers a high visibility of the quantum interference of the heralded entanglement between different anti-Stokes modes, as well as a tomography results of the joint entangled state of the two photonic qubits with high concurrence and fidelity. Furthermore, the broadband feature may enable future quantum networks being operated at high data rates, providing the accessibilities for overcoming the losses and inefficiency to generate and analyze the quantum entanglement. We also harness the memory-built-in feature to test delay-choice gedanken experiment, beside fundamental interest, which illustrates the capacities of quantum networks with shared quantum entanglement and quantum memory. 

The demonstrated low-noise, broadband and room-temperature building block is key to construct future scalable and environment-friendly quantum networks. Along this way, a few milestone works can be done for developing entirely new capacities of engineering quantum systems towards real-applications, while pushing the boundaries of quantum-classical transition. One may teleport arbitrary qubit to separated atomic ensembles consisted of motional atoms at ambient environment. One may build broadband and room-temperature quantum repeaters\cite{RMP_QR, Briegel_repeaters,nodes_QNet} capable of distributing quantum resources at high-speed fashion. The room-temperature atomic ensemble has the advantages of being cost-effective and easily miniaturized compared with other systems, which may exert great superiorities in building practical quantum networks, especially for the scenery of outer space. In addition, it is quite promising to prolong the lifetime of the heralded quantum entanglement among nodes, for instance, to preserve the coherence by the anti-relaxation coating in the vapor cell\cite{balaba_coating,polsik_DLCZ}, to transfer the spin wave of alkaline metal atoms to noble-gas nuclear spins in the regime of spin exchanging\cite{prl_nuclear spin,ofer_noble-gas}, etc. 

\subsection*{Acknowledgments.}
The authors thank Jian-Wei Pan for helpful discussions. This research was supported by the National Key R\&D Program of China (2019YFA0308700, 2017YFA0303700), the National Natural Science Foundation of China (61734005, 11761141014, 11690033), the Science and Technology Commission of Shanghai Municipality (STCSM) (17JC1400403), and the Shanghai Municipal Education Commission (SMEC) (2017-01-07-00-02- E00049). X.-M.J. acknowledges additional support from a Shanghai talent program.\\

\subsection*{Methods}

\paragraph*{Experimental details:} 
As the entire experimental setup shows in Figure 2, the two $^{133}$Cs cells, separated by 30 $cm$, are placed into a magnetic shielding and heated to 61$^{\circ}$C for getting a large optical depth. In order to alleviate the collisions between cesium atoms, we have injected 10 $Torr$ Ne buffer gas into the vapor cell. Benefit from the developed precise frequency locking system, the frequency of the optical control light can be locked to a fixed detuning and conveniently tuned on demand, which helps to create and verify the quantum entanglement precisely. The pump light, being resonated to the $\left | e \right \rangle \rightarrow \left | s \right \rangle$ transition, is directed from one port of Wollaston prism (WP), which propagates along the same path of the optical control light but with opposite direction and is employed to initialize the state of atoms. The creation of the pump and control light is generated in a programable fashion. To be specific, the optical control light pulse is $2ns$ generated by a high speed light modulator and the pump light pulse is $2us$ propagating along the diffraction path of an acousto-optical modulators (AOM).

The Stokes and anti-Stokes photons generated via spontaneous Raman scattering process are orthogonally polarized with the optical pump and probe light\cite{Nunn_Ph.D_2008}. We separate the control light and signal photons by their polarization via a high-extinction WP, which increases the signal-to-noise ratio for the Stokes and anti-Stokes photons. Besides the polarization filtering, we have built four sets of broadband Fabry-P\'erot cavities to extract the signal photons from the noise, whose single cavity can reach the transmission rate of $92\%$ and the extinction rate of $500:1$. In order to analyze the correlated photon pairs with individual modules (i.e. the heralding part and verifying part, as shown in Fig. 2), the Stokes and anti-Stokes photon are separated with their time sequences by applying a 100 $ns$ controlling signal to the AOM, in which the anti-Stokes photons pass through the way along the original incident direction and the Stokes photons propagate along the diffraction path.

\paragraph*{The phase locking for hetero-beam with orthogonal polarization:} 
The phase stabilization is necessary for faithful and stable observation of the interference of the heralded entanglement of different anti-Stokes modes. Either the phase $\varphi  _{S}$ or $\varphi  _{AS}$ contains two similar components, i.e. the phase difference resulting from the optical pump or probe pulse at two ensembles, and the phase difference accumulated from the propagation of Stokes or anti-Stokes photon. Due to the slowly drift of temperature or mechanical vibrations of optical devices at the ambient environment, the phases about the propagation of signal photons (the Stokes and anti-Stokes photons) will suffer from dramatic variations. In order to observe and verify the genuine entanglement between the two room-temperature ensembles from trial to trial, we should stabilize the interferometer loop in Fig. 2 at a fixed phase.

The auxiliary light field for phase locking of the interferometer loop in Fig. 2 has the same frequency as optical pump and probe pulse, but in the form of continuous wave. Due to the orthogonal polarization between the optical control light and scattering photons, it seems impossible to make the auxiliary light field propagating through the same path as the original interferometer loop, because the Glan-Taylor prism used for purifying the polarization of control light only allows one polarized light pass through, and blocks the orthogonal one with a extinction rate up to $10^5$. As the inset of Figure 2 shows, we employ a special half-wave plate with a hole in the center to change the polarization of auxiliary light field while keep the polarization of the optical control light unaffected. According to the interference results of the auxiliary field, the feedback electric controlling signal will be sent to the Piezoelectric ceramics to compensate the optical path.

\paragraph*{Quantum interference for the heralded entanglement of anti-Stokes modes:} 
For evaluating the coherence of the heralded entangled modes between the orthogonal polarization of anti-Stokes photon (can be written as $\frac{1}{\sqrt{2}}\left (\left | H \right \rangle_{L}^{AS}\pm e^{i\varphi  _{AS}}\left | V \right \rangle_{R}^{AS} \right )$), we need to control the phase $\varphi  _{AS}$ to acquire the interference results by means of the projection measurements (made up of the half-wave plate and polarizing beamsplitter shown in Fig. 2). The manipulation of this phase is achieved in the way of Pancharatnam-Berry's phase \cite{science_diamond,Berry_phase,Langford_Ph.D_2007}, which is realized by using two quarter-wave plates and a half-wave plate labelled by the component phase shifter in the verifying module shown in Fig. 2.

\clearpage

\begin{figure}[htbp]
	\centering
	\includegraphics[width=1.0\linewidth]{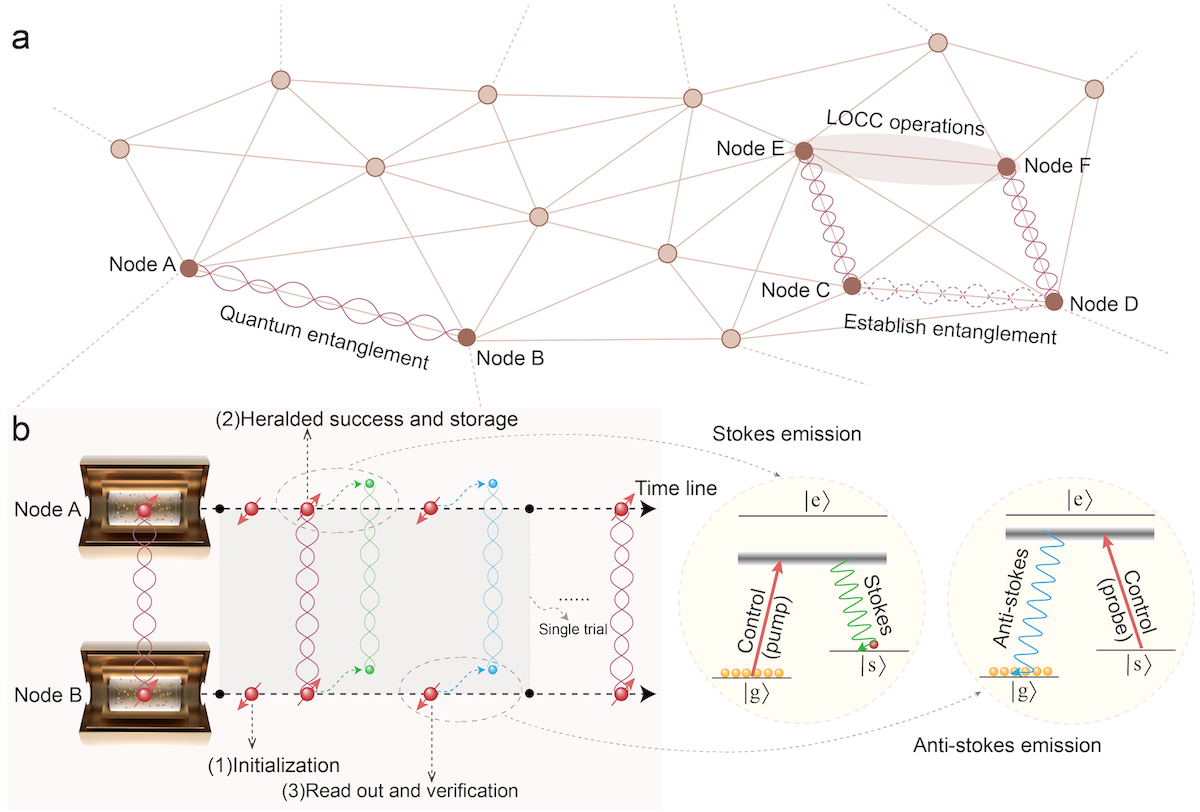}
	\caption{\textbf{Heralded Quantum Entanglement for Quantum Network at Room Temperature.} \textbf{a.} Sketch of quantum networks. The quantum resources can be disseminated through the quantum networks via the local operations and classical communications (LOCC) between different nodes. \textbf{b.} The time sequence of the generation and verification of the heralded quantum entanglement and the corresponding energy levels. The whole protocol contains three main stages, i.e. initializing, heralding and verifying processes, which form a single experimental trial. The atomic ensembles in each node has equal probability to contain a single excitation shared by all the atoms when the entanglement has been established, which is symbolized by spin-up in the graph. The atomic entanglement is heralded by the single Stokes photon which is also entangled between two different polarization modes. In the energy levels of the heralding and verifying processes, the solid lines represent three-level $\Lambda$-type configuration of atoms: the two ground states label $\left|g\right\rangle$ ($6S_{1/2}, F=3$) with electronic spin down and $\left|s\right\rangle$ ($6S_{1/2}, F=4$) with electronic spin up, which are hyperfine ground states of cesium atoms; the excited state labels $\left| {\rm{e}} \right\rangle $ ($6P_{3/2}, F^{'}=2,3,4,5$). The shaded area between energy levels represent broad virtual energy levels induced by the short optical pump and probe laser pulse.}
	\label{f2}
\end{figure}

\clearpage

\begin{figure}[htbp]
	\centering
	\includegraphics[width=1.0\linewidth]{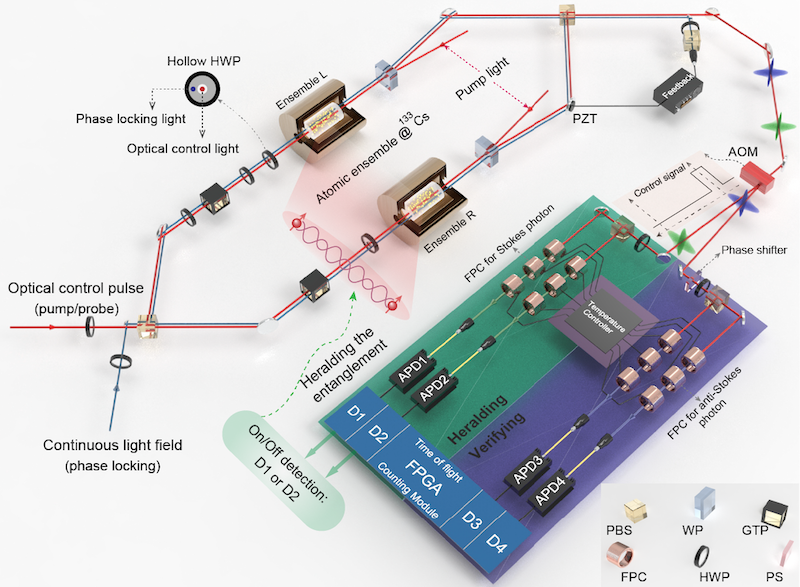}
	\caption{\textbf{Experimental setup.} The optical control pulse is directed into the Mach-Zehnder interferometer to generate the entanglement between the L and R ensembles. For combine photons from two ensembles at the PBS, the polarization of the control light in the left arm is transformed into being orthogonal to the right arm by the HWP after the GTP. The continuous auxiliary light field for phase locking is paralleled to the path of the control light, but has a small spatial shift to go thorough the hollow HWP for changing its polarization. This design is mainly due to the orthogonal polarizations between the control light passing through the GTP and the scattered signal photons passing the port of WP. The interference results of the auxiliary light field will be input to the processor, and then a feedback electric signal will be given to drive PZT to actively lock the phase of the interferometer. We utilize the time difference between the Stokes and anti-Stokes photons to separate their paths by applying a 100 $ns$ control signal to the AOM. By this way, the two photons are directed into different analyzing modules, i.e. the heralding part and verifying part. The PS is used to adjust the Pancharatnam-Berry's phase to acquire the interference results of the heralded anti-Stokes photon. PBS: polarization beam splitter, WP: Wollaston prism, GTP: Glan-Taylor prism, FPC: Fabry-P\'erot cavity, HWP: half wave plate, PS: phase shifter, PZT: piezoelectric ceramics, AOM: acousto-optical modulators, APD: avalanche photodiode detector. }
	\label{f2}
\end{figure}

\clearpage

\begin{figure}[htbp]
	\centering
	\includegraphics[width=1\linewidth]{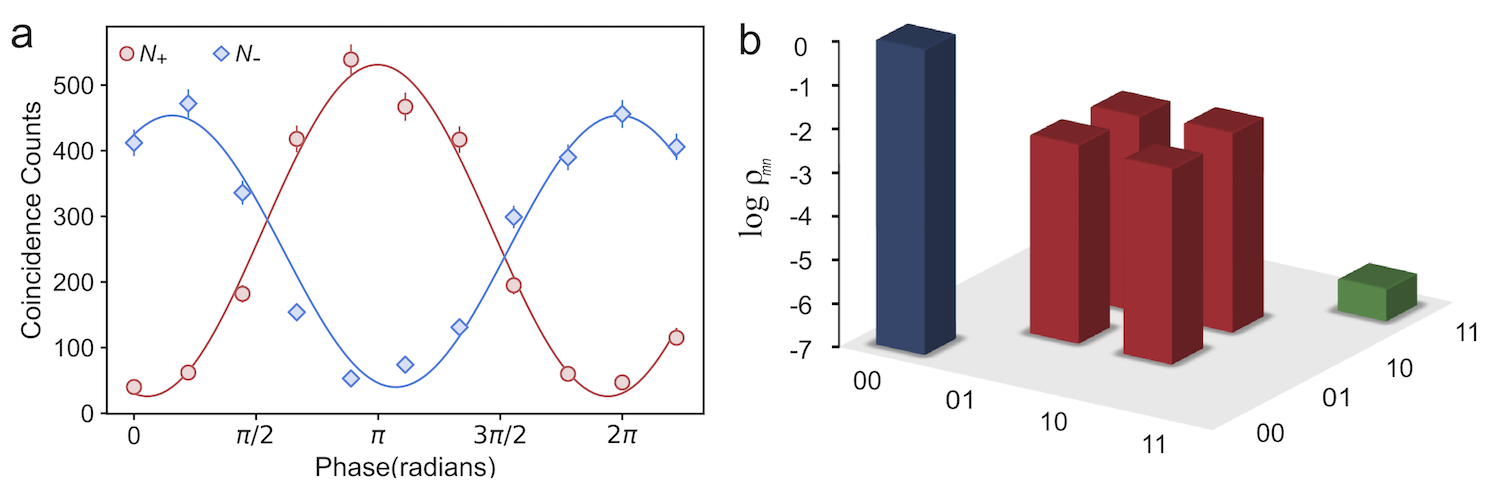}
	\caption{\textbf{The measured coherence between the two entangled atomic ensembles.} \textbf{a.} The coherence of the entanglement between the L and R ensembles is retrieved by the interference results of the mapped-out anti-Stokes photon, which is also heralded by the Stokes photon. The coincidence counts of $N_{\pm}$, stand for the registration counts of $D_{1}$ and $D_{3,4}$ respectively. The estimates of visibilities for $N_{\pm}$ is $V_{+}=(90\pm2)\%$ for $N_{+}$, and $V_{-}=(84\pm2)\%$ for $N_{-}$. The error bars are derived from the Poisson distribution of the finite coincidence counts.  \textbf{b.} The reconstructed density matrix of the heralded entanglement between different anti-Stokes modes. The density matrix elements are $p_{01}=3.1\times 10^{-3}$, $p_{10}=3.5\times 10^{-3}$, $d=V\times(p_{01}+p_{10})/2=2.9\times 10^{-3}$, $p_{11}=5.5 \pm1.1\times 10^{-7}$. The $p_{11}$ indicates the high-order excitation events, which has been much smaller than the events of single excitation, so we can neglect the higher-order excitation events contributing to the density matrix.}
	\label{f2}
\end{figure}

\clearpage

\begin{figure}[htbp]
	\centering
	\includegraphics[width=1\linewidth]{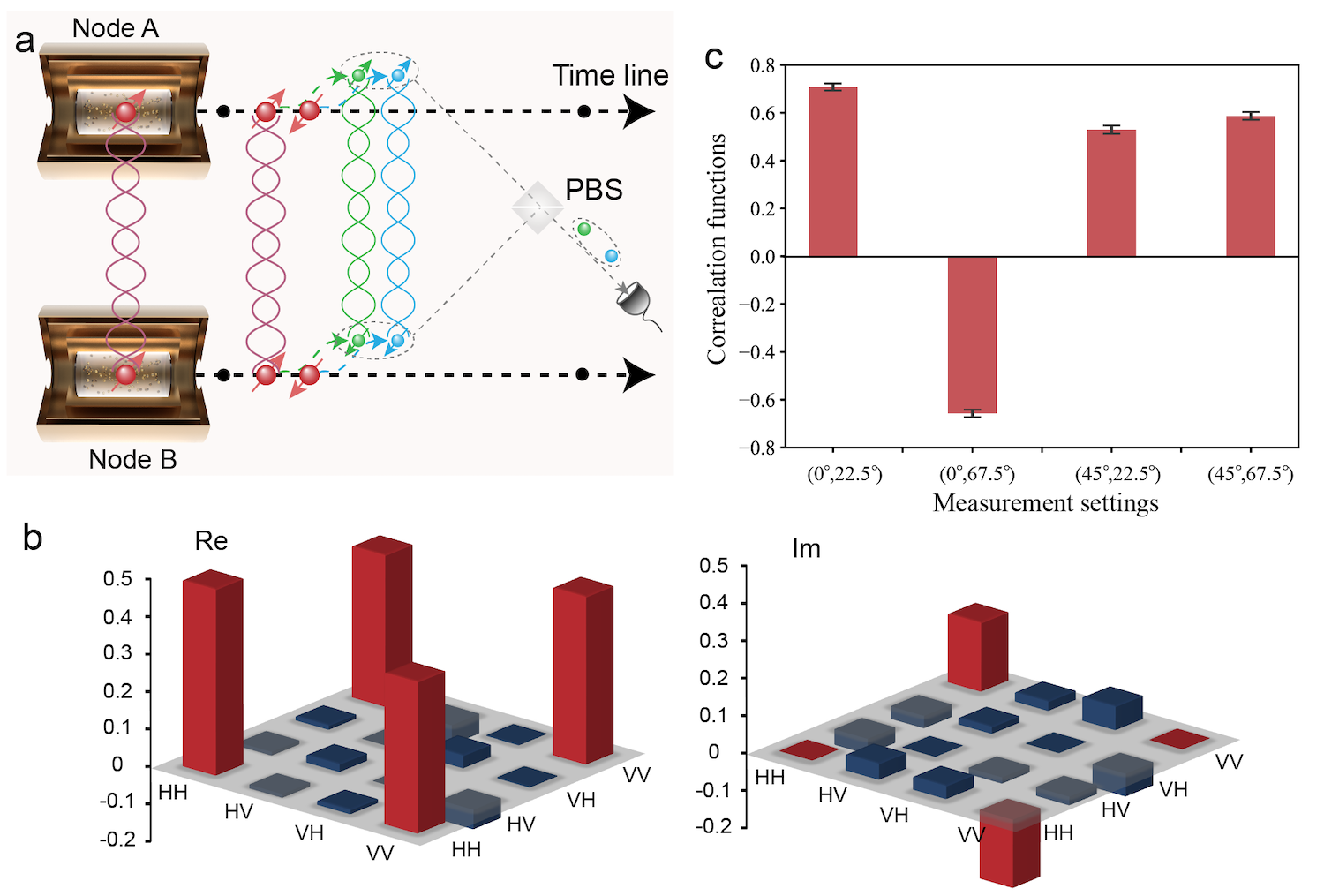}
	\caption{\textbf{The reconstructed density matrix and CHSH inequality test of the joint entangled state.}  \textbf{a.} The Stokes and anti-Stokes photons from a single atomic ensemble (L or R) have the same polarization, so the joint state of the photons pair can be written as $\left | H \right \rangle\left | H \right \rangle$ or $\left | V \right \rangle\left | V\right \rangle$. Considering the superposition of the two atomic ensembles, the whole state of the correlated photon pair is an entangled state of two photonic qubits. For reconstructing the density matrix of the quantum state by tomography, the polarization entangled state is projected to different tensor bases in different measurement settings. \textbf{b.} The real (Re) component and imaginary (Im) component of the density matrix. The subspaces in the reconstructed density matrix are labelled by $HH$, $HV$, $VH$ and $VV$. For instance, the $HH$ means that the polarizations of the joint Stokes photon and anti-Stokes photon are horizontally polarized, the others have the similar meanings. \textbf{c.} The correlation function values in different measurement settings in the CHSH inequality test. The error bars come from the Poisson statistics of the coincidence counts. }
	\label{f4}
\end{figure}

\clearpage

\begin{figure}[htbp]
	\centering
	\includegraphics[width=1\linewidth]{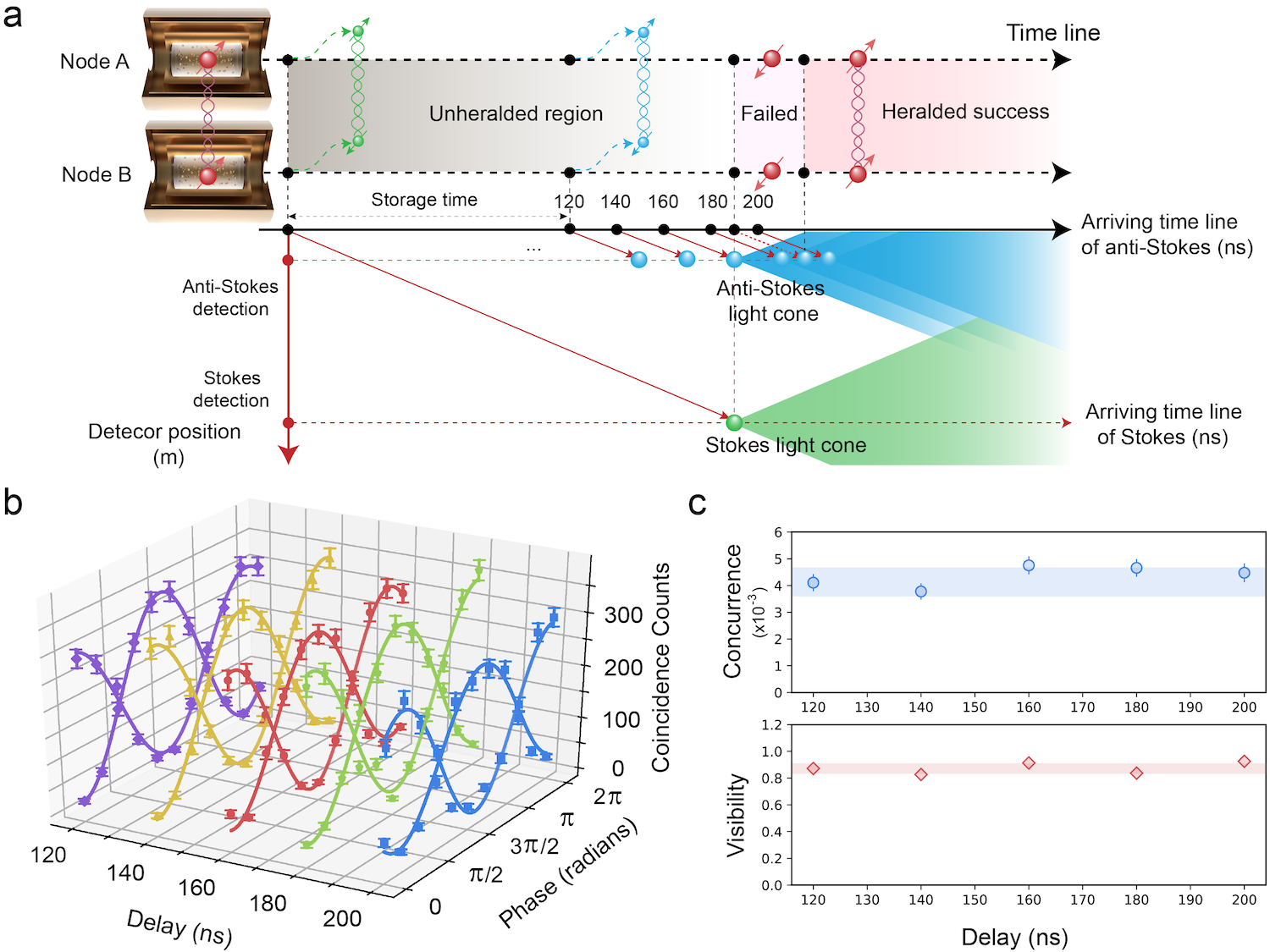}
	\caption{\textbf{Delay-choice gedanken experiment with heralded quantum entanglement and built-in quantum memories.}  \textbf{a.} The space-time diagram of delay-choice test. The projection measurement of Stokes mode has been delayed by $160ns$, so we can manipulate the storage lifetime of the heralded entanglement to change the detection choices. The different detection situations of the Stokes and anti-Stokes photon mean that the totally distinguished heralded cases of the atomic ensembles. The forward light cones of the detections of Stokes and anti-Stokes photon do not overlap, which means that they are independent from each other and do not have the causal correlations. \textbf{b.} The coherence results of the entangled anti-Stokes modes at different retrieval delays. The error bars come from the Poisson statistics of the coincidence counts. \textbf{c.} The variations of visibilities and concurrences with the delays. With an approximate estimation, the visibility is nearly a constant at $V=(87.5\pm4)\%$, and the concurrence is approximately constant with an average of $(4.35\pm0.36)\times 10^{-3}$.}
	\label{f4}
\end{figure}

\clearpage

\end{document}